\documentstyle[prl,aps]{revtex}
\input BoxedEPS.tex
\SetOzTeXEPSFSpecial
\HideDisplacementBoxes

\begin{document}

\twocolumn[\hsize\textwidth\columnwidth\hsize
           \csname @twocolumnfalse\endcsname
\title{Electrical transport in the ferromagnetic state of manganites:
Small-polaron metallic conduction at low temperatures}
\author{G. M.  Zhao$^{(1,2)}$, V. Smolyaninova $^{(1)}$, W.
Prellier$^{(1,*)}$ and H. Keller $^{(2)}$}

\vspace{1cm}
\address{$^{(1)}$Center for Superconductivity Research,
Physics Department, University of Maryland, College Park, MD 20742,
USA\\
$^{(2)}$ Physik-Institut der Universit\"at Z\"urich, CH-8057
Z\"urich, Switzerland}

\maketitle
\noindent
\begin{abstract}
We report measurements of the resistivity in the ferromagnetic state of
epitaxial thin films of
La$_{1-x}$Ca$_{x}$MnO$_{3}$ and the low temperature specific heat of
a polycrystalline La$_{0.8}$Ca$_{0.2}$MnO$_{3}$. The resistivity below
100 K can be well fitted by $\rho - \rho_{o} =
E\omega_{s}/\sinh^{2}(\hbar\omega_{s}/2k_{B}T)$ with
$\hbar\omega_{s}/k_{B}$ $\simeq$ 100 K and $E$ being a constant. Such behavior
is consistent with small-polaron coherent motion which involves
a relaxation due to a soft optical phonon mode. The specific heat data
also suggest the existence of such a phonon mode. The present results thus
provide
evidence for small-polaron metallic conduction in the ferromagnetic
state of manganites.

\end{abstract}

\vspace{1cm}

]
\narrowtext
Nearly a half century ago, Volger \cite{Volger} first observed a
large magnetoresistance in a bulk sample of the manganite
La$_{0.8}$Sr$_{0.2}$MnO$_{3}$ near room temperature. The recent
discovery of ``colossal" magnetoresistance (CMR) in thin films of
Re$_{1-x}$A$_{x}$MnO$_{3}$ (Re = a rare-earth ion, and A = a divalent
ion) \cite{Von} has attracted renewed interest in these systems.
In order to understand the microscopic origin
of the CMR effect, extensive studies of magnetic, structural and
transport properties have been carried out on these
materials \cite{Art}. The physics of manganites
has primarily been
described by the
double-exchange model \cite{Zener}. Recent
calculations \cite{Millis1,Alex} show
that a second mechanism such as a strong polaronic effect
should be involved to explain the basic physics. Many
recent experiments have provided compelling evidence for the existence
of polaronic charge carriers in the
paramagentic state of manganites
\cite{Jaime}.

However, the electrical transport mechanism
below T$_{C}$ is poorly understood. At low temperatures, a dominant
$T^{2}$ contribution in resistivity is generally observed, and has
been ascribed to electron-electron scattering \cite{Urushibara}.
Jaime {\em et al.} \cite{Jaime2} have
recently shown that the resistivity is essentially temperature
independent below 20 K and exhibits a strong $T^{2}$ dependence above
50 K. In addition, the coefficient of the $T^{2}$ term is about 60
times larger than that expected for electron-electron scattering. They
thus ruled out the electron-electron scattering as the conduction
mechanism and proposed single magnon scattering with a cutoff at
long wavelengths. Their scenario can qualitatively explain the
observed data, but there is no quantitative agreement between the calculated
and experimental results.

In this letter, we report measurements of the resistivity in the
ferromagnetic state of epitaxial thin films of
La$_{1-x}$Ca$_{x}$MnO$_{3}$ and the low temperature specific heat of
polycrystalline La$_{0.8}$Ca$_{0.2}$MnO$_{3}$. The resistivity below
100 K obeys a formula $\rho - \rho_{o} =
E\omega_{s}/\sinh^{2}(\hbar\omega_{s}/2k_{B}T)$ with
$\hbar\omega_{s}/k_{B}$ $\simeq$ 100 K.
Such behavior
is consistent with small-polaron coherent motion which involves
a relaxation due to a low-lying optical phonon mode.

\begin{figure}[htb]
    \ForceWidth{6.6cm}
	\centerline{\BoxedEPSF{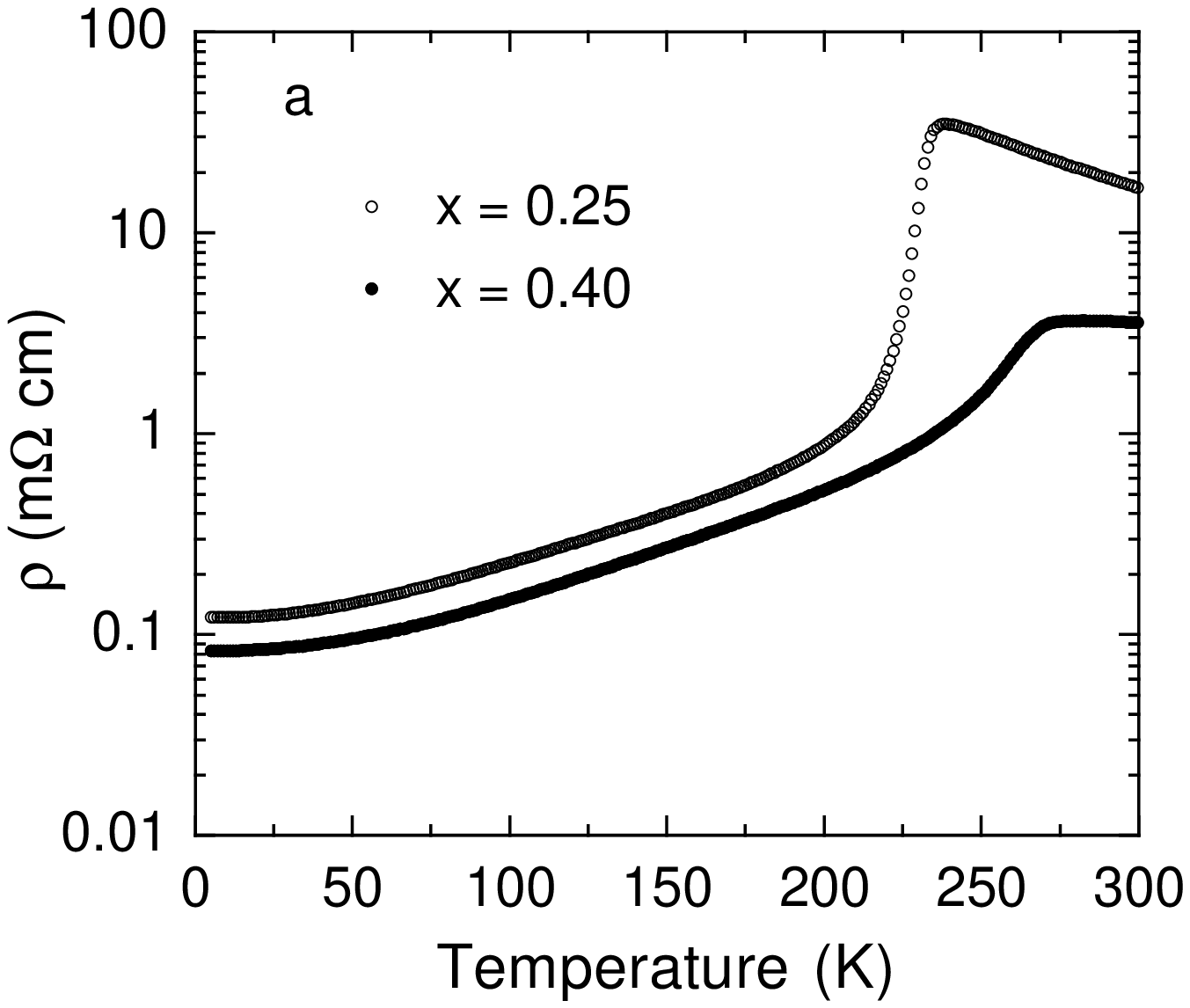}}
	\ForceWidth{6.6cm}
	\centerline{\BoxedEPSF{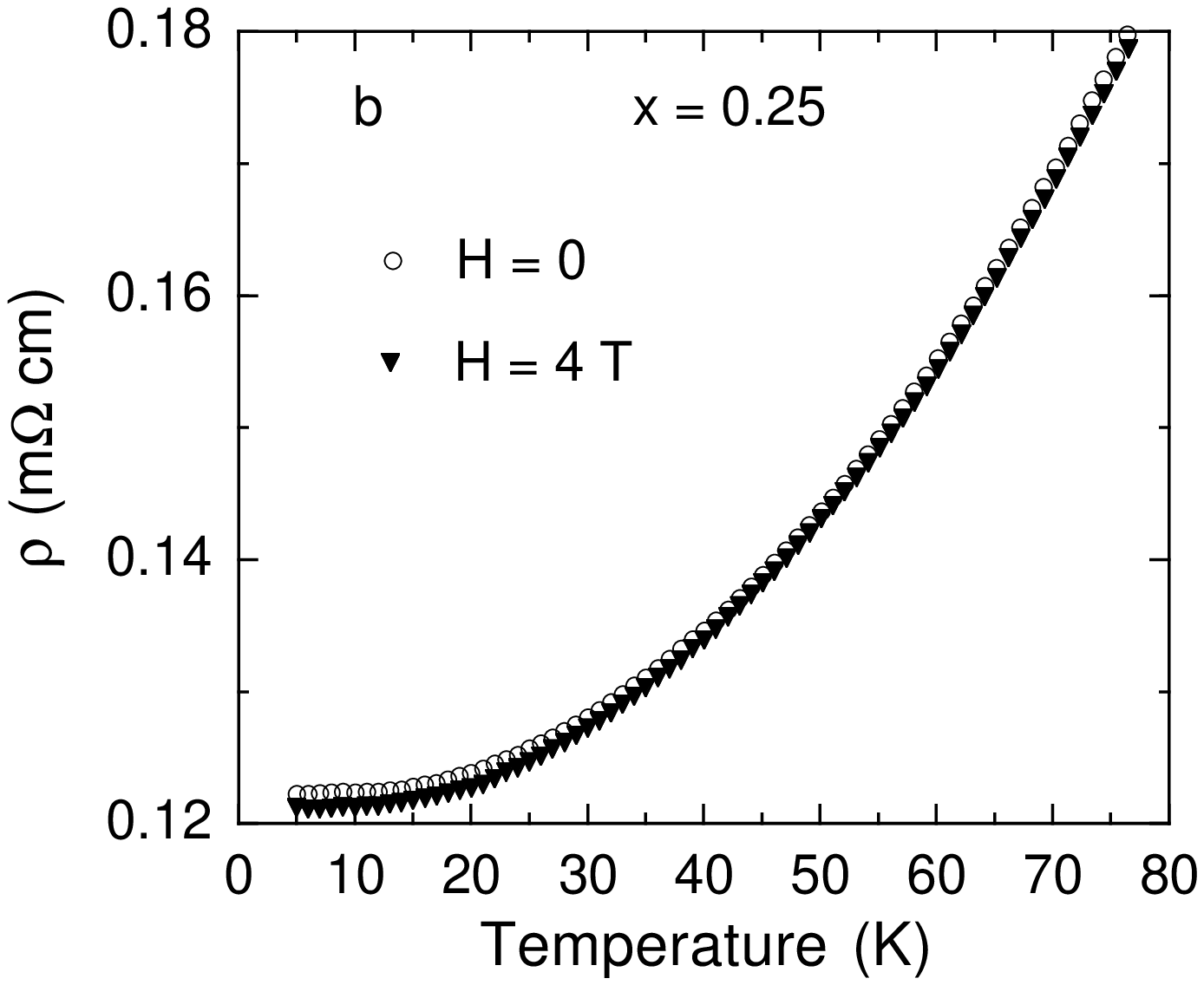}}
	\vspace{0.2cm}
	\caption[~]{(a) The zero-field resistivity $\rho(T)$ of
the thin films La$_{1-x}$Ca$_{x}$MnO$_{3}$ with $x$ =
0.25 and 0.40; (b) Low-temperature resistivity $\rho(T)$ of the $x$ = 0.25
film in zero and 4 Tesla magnetic field.}
	\protect\label{Fig.1}
\end{figure}

The epitaxial thin films of La$_{1-x}$Ca$_{x}$MnO$_{3}$ with $x$ =
0.25 and 0.40 were grown
on $<$100$>$ LaAlO$_{3}$ single crystal substrates by pulsed laser
deposition using a KrF excimer laser \cite{Prellier}. The deposition frequency
is 10 Hz and the laser energy density is about 1.5 J/cm$^{2}$. The films
were finally annealed
for 10 h at about 940 $^{\circ}$C and oxygen pressure of about 1 bar.
The thickness of the films are about 150 nm.
The polycrystalline sample of La$_{0.8}$Ca$_{0.2}$MnO$_{3}$ was
prepared by
conventional solid state reaction using dried La$_{2}$O$_{3}$, MnO$_{2}$
and CaCO$_{3}$ \cite{Zhao99}. The resistivity was measured using the van
der Pauw technique, and the
contacts were made by silver paste. The absolute inaccuracy of the
resistivity is less than 5$\%$.
The measurements were carried out from 5 to 380 K in a Quantum Design
measuring system. The specific heat was measured in a temperature
range of 2-16 K by relaxation calorimetry with an absolute inaccuracy of
10$\%$.


Fig.~1a shows the zero field resistivity of
the thin films La$_{1-x}$Ca$_{x}$MnO$_{3}$ with $x$ =
0.25 and 0.40, respectively. There are metal-insulator
transitions at about 240 K and 280 K for $x$ = 0.25 and 0.40,
respectively. The residual resistivity is 123
$\mu\Omega$cm for $x$ = 0.25, and 84 $\mu\Omega$cm for $x$ = 0.40.
The values of the residual resistivity in these films are even smaller
than that for single crystalline samples \cite{Snyder}. This indicates that
the quality of
the films is high, which allows one to study the intrinsic electrical
transport
properties of this system.

\begin{figure}[htb]
    \ForceWidth{6.5cm}
	\centerline{\BoxedEPSF{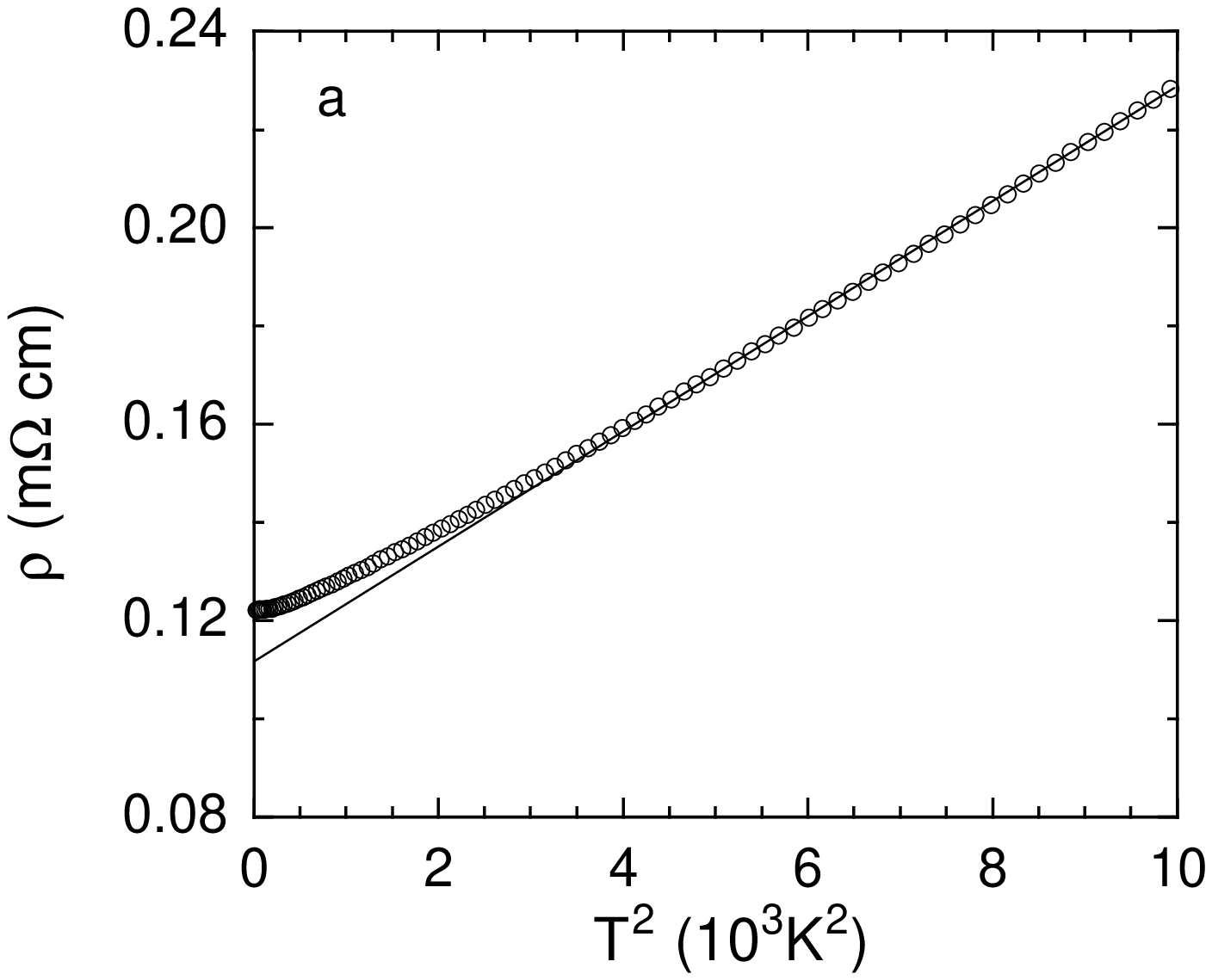}}
	\ForceWidth{6.5cm}
	\centerline{\BoxedEPSF{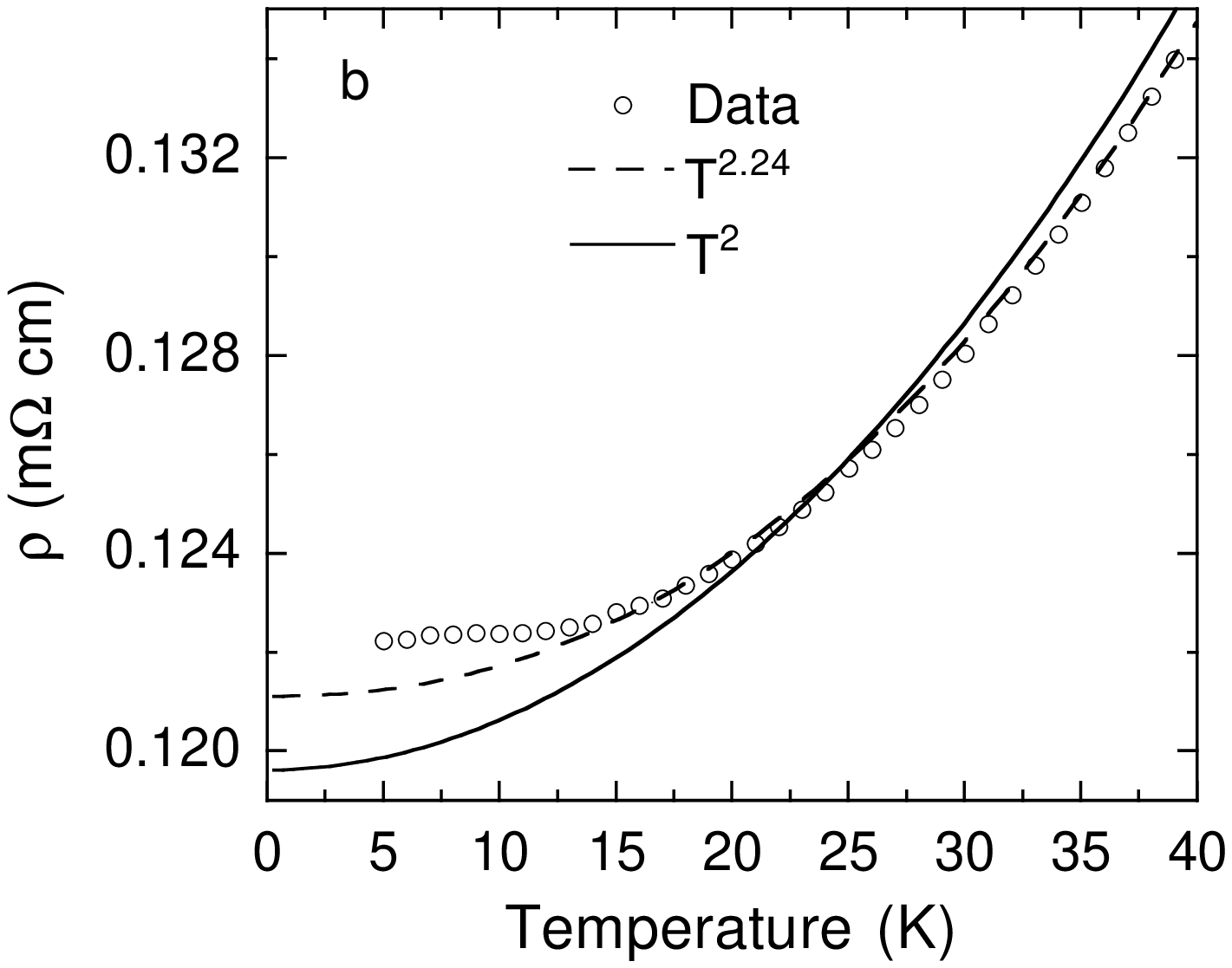}}
	\vspace{0.2cm}
	\caption[~]{(a) Resistivity $\rho(T)$ vs
$T^{2}$ for $x$ = 0.25. (b) Resistivity $\rho(T)$ vs $T$ for $x$ = 0.25.
The solid and dash lines are the curves for the best $T^{2}$ and
power-law fits to the data below 80
K, respectively.}
	\protect\label{Fig.2}
\end{figure}
In Fig.~1b we plot the low-temperature resistivity of the $x$ = 0.25
film in zero and 4 Tesla magnetic field. Basically, there is a
negligible magnetoresistance effect below 80 K, in agreement with
Ref.~\cite{Jaime2}.

In order to see more clearly whether the low temperature resistivity
has a $T^{2}$ contribution, we show, in Fig. 2a, $\rho(T)$ vs $T^{2}$
for the $x$ = 0.25 film.
It is apparent that the resistivity exhibits a dominant quadratic temperature
dependence above 60 K, in agreement with Ref. \cite{Jaime2}. We
try to fit the data below 80 K by $\rho (T) = \rho_{0} + FT^{2}$ and by
a power-law $\rho (T) = \rho_{o} + FT^{n}$. Both fits are quite bad as
seen clearly from Fig.~2b where only the data below 40 K are shown.
Even the best power-law fit with $n$=2.24 deviates from the data
substantially below 20 K where the resistivity is nearly temperature
independent.

Alternatively, one should consider a contribution from electron-phonon
scattering. At low temperatures, the acoustic phonon
scattering would give a $T^{5}$ dependence, which is not consistent
with the data. Recently, Alexandrov and Bratkovsky \cite{Alex} have proposed a
theory for colossal magnetoresistance in doped manganites. Their model
predicts that polaronic transport is the prevalent conduction mechanism
even below $T_{C}$. This has been partially supported by the low
temperature optical data which reveal a small coherent Drude weight and
a broad incoherent spectral feature \cite{Simpson,Lee}. If their model is
relevant, the
temperature dependence of the resistivity at low temperatures should
be consistent with small polaron transport.

Although a theory of small polaron conduction at low temperatures was
worked out more than 30 years ago \cite{Lang}, no experimental data have been
used to compare with the theoretical prediction. The theory shows
that \cite{Lang}, for $k_{B}T < 2t_{p}$, the resistivity is given by
\begin{equation}\label{eq1}
\rho(T) =
(\hbar^{2}/ne^{2}a^{2}t_{p})(1/\tau),
\end{equation}
where $t_{p}$ is the hopping
integral of polarons, $n$ is the carrier density, $a$ is the lattice
constant, and $1/\tau$ is
the relaxation rate:
\begin{equation}\label{eq2}
1/\tau =
\sum_{\alpha}A_{\alpha}\omega_{\alpha}/\sinh^{2}(\hbar\omega_{\alpha}/2k_{B}T),
\end{equation}
where $\omega_{\alpha}$ is the average frequency
of one optical phonon mode, $A_{\alpha}$ is a constant, depending on the
bare conduction bandwidth and the
electron-phonon coupling
strength. It is worth noting that the above expression
for $1/\tau$ has been generalized from one optical phonon mode to multiple
modes since complex compounds such as manganites contain several
optical phonon modes. From the above equations, one can see that
only the soft modes may contribute to the resistivity at low temperatures
due to the factor of $1/\sinh^{2}(\hbar\omega_{\alpha}/2k_{B}T)$.
By inclusion of impurity scattering, the total resistivity at low
temperatures is
\begin{equation}\label{eq}
\rho(T) = \rho_{0} + E\omega_{s}/\sinh^{2}(\hbar\omega_{s}/2k_{B}T),
\end{equation}
where $\omega_{s}$ is the average frequency of the softest optical
mode, and $E$ is a constant, being proportional to the effective mass of
polarons. Eq.~3 is valid only if all other optical modes that
are strongly coupled to the carriers have much higher frequencies than
the softest mode. Otherwise, one has to use a more general formula
that includes the contributions from all the modes.

In Fig. 3, we show the low temperature resistivity $\rho(T)$ for $x$ = 0.25
and
0.40 films. The data can be fitted by
Eq.~3 with $\hbar\omega_{s}/k_{B}$ = 86(2) K for $x$ = 0.25, and 101(2) K for
$x$ = 0.40. This indicates that a soft mode with $\hbar\omega_{s}/k_{B}$ of
about 100
K has a strong coupling with the carriers and thus contributes to the
scattering. Such a soft mode is commonly present in perovskite oxides
such as cuprates and
nickelates. For La$_{1.85}$Sr$_{0.15}$CuO$_{4}$, a soft mode with
$\hbar\omega_{s}/k_{B}$ of about 115 K has been observed from both
neutron scattering and specific heat measurements \cite{APR}.

In order to check whether there exists this soft mode, we measured
the low temperature specific heat of a polycrystalline sample of
La$_{0.8}$Ca$_{0.2}$MnO$_{3}$, as shown in Fig.~4a. The specific heat
in this temperature
region can be expressed as
\begin{equation}\label{c}
C(T) = BT^{1.5}+\gamma T+\beta T^{3}+ A/T^{2} + \Delta C(T),
\end{equation}
where the first three terms arise from magnons, charge carriers, and
acoustic phonons, respectively, the fourth term is from a
Schottky anomaly, the last term $\Delta C(T)$ =
$D(\hbar\omega_{s}/k_{B}T)^{2}\exp(\hbar\omega_{s}/k_{B}T)/
[\exp(\hbar\omega_{s}/k_{B}T)-1]^{2}$, which is contributed from
an optical mode. The solid line is the best fit to the data with
four fitting parameters: $\gamma$ = 7.17(6) mJ/mole K$^{2}$, $\beta$ =
0.200(2) mJ/mole K$^{4}$, $D$ = 3.30(8) J/mole K and
$\hbar\omega_{s}/k_{B}$ = 95.9(7) K, and with two fixed parameters:
$A$ = 8.0 mJ K/mole \cite{Gordon} and $B$ = 0.4 mJ/mole K$^{2.5}$
\cite{note}, The $D$ value obtained for the
manganite is nearly the same as that ($\sim$ 3.5 J/mole K) for
La$_{1.85}$Sr$_{0.15}$CuO$_{4}$ \cite{APR}. This implies that the soft
mode in the manganite is similar to that in the cuprate. If we fix $D$ =
$R/2$ = 4.15 J/mole K, the quality of the fit remains unchanged, and
the value of $\hbar\omega_{s}/k_{B}$ increases to 104 K. In order to
see more clearly the contribution due to the optical mode, we
plot $\Delta C(T)/T$ vs T$^{2}$ in Fig.~4b. It is apparent that
the data can be well fitted by an Einstein mode with
$\hbar\omega_{s}/k_{B}$ of about 100
K. This justifies the above explanation to the low temperature
resistivity data.

\begin{figure}[htb]
    \ForceWidth{6.6cm}
	\centerline{\BoxedEPSF{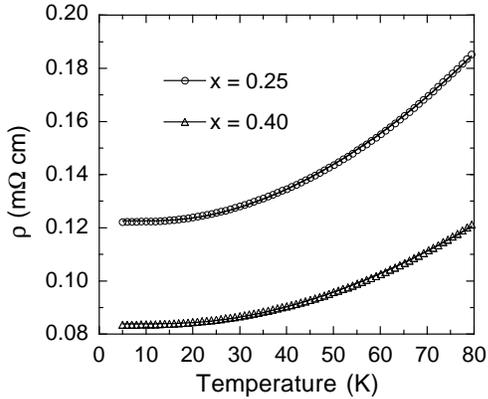}}
	\vspace{0.2cm}
	\caption[~]{Low temperature resistivity $\rho(T)$ for $x$ = 0.25 and
0.40 films. The solid lines are fitted curves by Eq.~3 with
$\hbar\omega_{s}/k_{B}$ = 86(2) K for $x$ = 0.25, and 101(2) K for
$x$ = 0.40.}
	\protect\label{Fig.3}
\end{figure}
Such a soft mode should also be observed by other
experimental techniques such as Raman scattering.
Raman spectra of a La$_{0.7}$Ca$_{0.3}$MnO$_{3}$ thin film \cite{Pod}
reveals a phonon mode with a
frequency $\hbar\omega_{s}/k_{B}$ = 127 K, which is close to
that (101 K) deduced from the resistivity data of
La$_{0.6}$Ca$_{0.4}$MnO$_{3}$. This is reasonable since the frequencies
of the zone-center (Raman active) optical modes are normally
higher than the average ones. The other modes with
$\hbar\omega/k_{B}$ = 599K, 686 K, and 964 K, respectively,
should have a negligible contribution to the scattering rate below 100
K due to the factor of $1/\sinh^{2}(\hbar\omega_{\alpha}/2k_{B}T)$ in
Eq.~2. But the mode with $\hbar\omega/k_{B}$ = 212 K may
contribute a little to the scattering rate above 50 K if the mode is
strongly coupled to the carriers. The fact that an excellent fit has been
achieved
without including this mode implies that the mode may have a weak
coupling to the carriers.
\begin{figure}[htb]
    \ForceWidth{6.6cm}
	\centerline{\BoxedEPSF{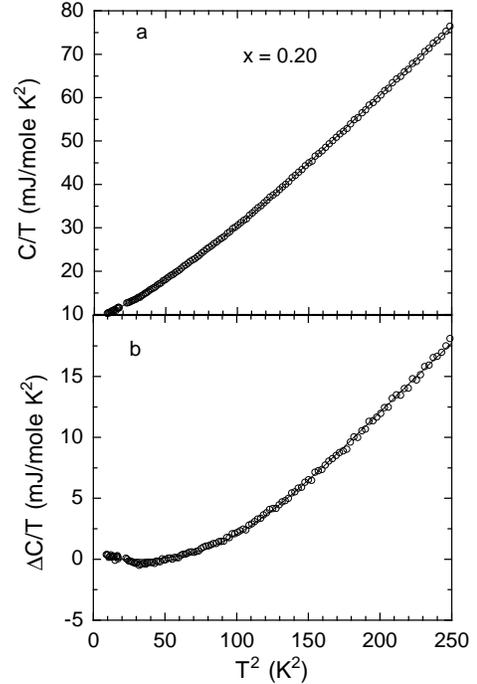}}
	\vspace{0.2cm}
	\caption[~]{(a) Low temperature specific heat of a polycrystalline
sample of
La$_{0.8}$Ca$_{0.2}$MnO$_{3}$; (b) Specific heat contributed from an
optical mode.}
	\protect\label{Fig.4}
\end{figure}

If the charge carriers at low temperatures are indeed of small
polarons, the effective mass of the carriers should be substantially
enhanced. It is possible to estimate the mass enhancement factor $f_{p}$
from the measured
screened plasma frequency $\Omega_{p}^{s}$, high-frequency dielectric constant
$\epsilon_{\infty}$, and effective plasma frequency $\Omega_{p}^{*}$. For
$x$ = 0.3 and 0.4, the screened plasma
frequencies are nearly the same and equal to 1.5 eV \cite{Lee,Saitoh99}.
This is consistent with
electron-energy-loss spectra (EELS) which show a maximum pre-O$_{K}$
peak intensity (related to doped hole density) around $x$ = 0.3
\cite{Ju}.
Then the bare plasma
frequency for $x$ = 0.3 is given by $\hbar\Omega_{p}^{b}$ =
$\sqrt{\epsilon_{\infty}}\hbar\Omega_{p}^{s}$ = 3.35 eV (we take
$\epsilon_{\infty}$ = 5.0 \cite{Kim1,Alexcond}).
So we obtain $f_{p}$ = 9 using the measured $\hbar\Omega_{p}^{*}$ =
1.1 eV \cite{Simpson}. The observed $\hbar\Omega_{p}^{b}$ is much larger
than that (1.9 eV) calculated from the local density approximation
(LDA) \cite{Pick}. This is quite reasonable since the LDA calculation does not
take into account a large onsite U on Mn sites \cite{Saitoh}. We can
also estimate the mass
enhancement factor from the specific heat data. Since the manganites are
doped charge-transfer insulators \cite{Ju,Saitoh}, the carrier
concentration for a doped manganite should be equal to $x$ per
cell when $x$$\leq$ 0.3 (see the argument of Ref. \cite{Ju}).
Then, the bare mass of the carriers is estimated to be
0.61 $m_{e}$ from the measured $\hbar\Omega_{p}^{b}$ for $x$ = 0.3, where
$m_{e}$ is the mass of
an electron. This leads to the bare electronic
specific heat coefficient $\gamma_{b}$ = 0.52 mJ/mole K$^{2}$ for $x$ = 0.2.
So $f_{p}$ = $\gamma$/$\gamma_{b}$ = 14 for $x$ = 0.20 where $\gamma$ =
7.17(6) mJ/mole K$^{2}$. Thus the mass enhancement factor is
substantial and typical for small Fr\"ohlich polarons
\cite{Alexcond,Alex99}. The small Fr\"ohlich polarons are not very heavy
because such polarons have a small size of the wavefunction but a large
size of lattice deformation \cite{Alex99}.

In summary, our low-temperature resistivity data on high-quality
epitaxial thin films of La$_{1-x}$Ca$_{x}$MnO$_{3}$ can be well
explained by a theory of small polaron metallic conduction which involves a
relaxation due to a soft optical phonon mode. This optical phonon mode has
a frequency of about 100 K, as revealed from both the resistivity and
specific heat data. Our present results provide compelling evidence for the
existence of
polaronic carriers in the low-temperature ferromagnetic state of
manganites, and support a CMR theory recently proposed \cite{Alex}.

{\bf Acknowlegement}: We would like to thank A. S. Alexandrov and A. M.
Bratkovsky, R. L. Greene, and A. Biswas for useful discussions.
The work was supported by the
American NSF MRSEC on Oxides and Swiss National Science Foundation.
~\\
~\\
*Present address: Laboratoire CRISMAT-ISMRA,14050 CAEN Cedex, France.

\bibliographystyle{prsty}

\begin{thebibliography}{10}
\bibitem{note} The value of $B$ is determined from comparing the
specific heat data of a single crystal La$_{0.7}$Sr$_{0.3}$MnO$_{3}$
in zero and 8.5 T magnetic field.
\end{thebibliography}


\end{document}